# Nanoscale Lasers Based on Carbon Peapods

*Shaolong He[*, †], Jianqi Shen[‡], Haiyang Li[†], and Hongnian Li[†]*

Department of Physics, Zhejiang University, Hangzhou 310027,

People's Republic of China

Center for Optical and Electromagnetic Research, Zhejiang University, Hangzhou 310027,

People's Republic of China

† Department of Physics, Zhejiang University, Hangzhou 310027, People's Republic of China

‡ Center for Optical and Electromagnetic Research, Zhejiang University, Hangzhou 310027, People's Republic of China

**ABSTRACT**: A scheme of nanoscale lasers based on the so-called carbon peapods is examined in details. Since there is considerable cylindrical empty space in the middle of a single-wall carbon nanotube (SWCNT), it can serve as a laser resonant cavity that consists of two highly reflecting, alignment "mirrors" separated by some distance. These mirrors refer to the ordered arrays of $C_{60}$ inside SWCNTs, which have photonic bandgap structures. Meanwhile, ideally single-mode lasers are supposed to be produced in the nanoscale resonant cavity.

One useful thing that comes out of the discovery of $C_{60}$ [1] may be new ideas and paths of investigation.[2-4] Carbon nanotubes (CNTs),[5] for example, are the direct result of fullerens research. It has been suggested that atoms and molecules can be trapped inside a SWCNT. Recently a supramolecular assembly comprising $C_{60}$ molecules inside single-wall carbon nanotubes ($C_{60}$@SWCNTs) known as carbon "peapods" was discovered by Smith et al. using transmission electron microscopy (TEM),[6] and was intensively studied both experimentally and theoretically.[7] These so-called carbon peapods not only represent a new family of nanostructured carbon, but also support the general idea that SWCNTs are ready to accept various types of compounds and further confirm the notion of being an ideal system for nanodevices. It is believed that the potential application of the carbon peapod is such that it can be utilized as a memory device capable of storing a bit of information by the voltage-driven shuttling of interior $C_{60}$ between the two endcaps of the single-wall carbon nanotubes. [8-9]

---

[*] Corresponding author. E-mail: shephy@zju.edu.cn

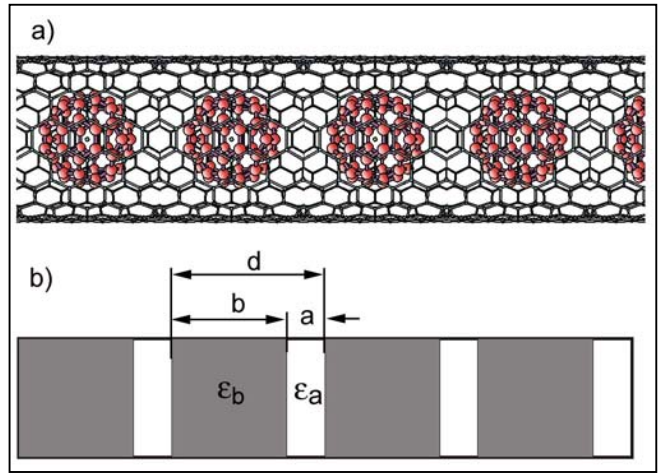

**Figure 1.** : Sketch of (a) a carbon peapod photonic crystal; (b) a one-dimensional photonic crystal.

More recently, ordered phase of fullerenes in CNTs over distances of 20-50 nm was observed. [10] Here, we present that such structure, where $C_{60}$-$C_{60}$ separations are slightly varied, is actually a one-dimensional photonic crystal (PC), [11,12] due to that the dielectric constant is periodically modulated. A sketch of the carbon peapod photonic crystal (CPPC) is shown in Figure 1(a). PCs have electromagnetic band gap, which overlaps the electronic band edge, and spontaneous emission can be rigorously forbidden. [11] To our best knowledge, there is no report on nanostructured photoic crystals. Therefore, CPPCs deserve detailed investigations of both fundamental interests and potential nanodevices applications.

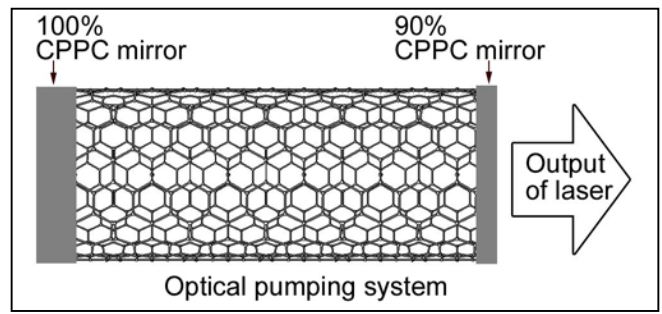

**Figure 2.** : Schematic plan of a nanoscale laser based on CPPCs, showing elements responsible for energy input, amplification, and output. The active atoms are capsulated inside the SWCNT and able to produce ideally single-mode lasers.

As a particular application, CPPCs can be employed as two perfect dielectric mirrors of a nanoscale laser's resonator, which is shown in Figure 2. Before we discuss the stability of the resonator as well as selecting gain medium, we would like to demonstrate the origin of the high reflectivity of CPPCs within some specified frequency band.

While calculations of band structures of three-dimensional PCs cannot be achieved without difficulty in mathematics, [13] analyses of one-dimensional PCs is much simpler using a standard transfer matrix method. [14] The calculated forbidden frequency for the CPPC shown in Figure 1 is

$$\omega_m = \frac{m\pi c}{a\sqrt{\varepsilon_a} + b\sqrt{\varepsilon_b}}, \quad (1)$$

in which $c$ is the speed of light in vacuum, and $\varepsilon_a$ ($\varepsilon_b$) the dielectric constant of media a (b). For a typical CPPC, $a = 0.3$ nm, $b = 0.7$ nm, [10,15] and $\varepsilon_b$ may be taken to be about 22; [16] thus we expect the lowest forbidden frequency for a CPPC to be about $2.63 \times 10^{17}$ Hz ($\lambda \approx 7.2$ nm). The calculated result reveals that x-rays, which have frequencies within the photonic band gap of CPPCs, could be trapped inside SWCNTs. Since carbon peapods can be intercalated with metal atoms [17] and metallofulleren can be inserted into SWCNTs [*e.g.*, (Gd@$C_{82}$)@SWCNTs], [18] the parameters of CPPCs can be conveniently changed in order to obtain proper bandgap structures of interest. This immediately suggests that CPPCs can be employed as perfect dielectric mirrors of the nanoscale lasers. Moreover optoelectronic circuits based on a CPPC can be easily accomplished by voltage driving [8,9] $C_{60}$ molecules inside the SWNT and therefore changing CPPC's parameter.

However, we are interested in of carbon peapods of finite length. Without going further than the transfer matrix method, one gets the reflectance and transmissivity for a CPPC consisting of $N$ segments

$$R_N = \left(\frac{\sin^2(N\delta)}{\delta}\right)(T^{-1} - 1),$$

$$T_N = 1 + \left(\frac{\sin^2(N\delta)}{\delta}\right)(T^{-1} - 1), \quad (2)$$

in which $\delta$ denotes the Bloch phase and $T$ the transmissivity for a unit cell. Equation (2) indicates that $R_N$, as well as $T_N$, depends on the number of the segments, namely the length of carbon peapods. Despite the fact that no one has yet rigorously quantified the filling of SWNTs with $C_{60}$ molecules, we can, in principle, obtain CPPC mirrors with reflectance near or much less than 100% by properly arranging the $C_{60}$ molecules inside the SWCNT. And they can be employed as total reflection and transparency mirror, respectively, of the nano-laser resonator.

One of the simplest but most important questions concerning a laser resonator is whether it is stable. If the ray remains within the resonant cavity after a sufficiently large number of reflections, it is said to be stable. In general, a stability criterion for a laser resonator can be expressed in terms of the radii of curvature of the mirrors and the distance separating the mirrors. In the case of CPPC mirrors, the radii of $C_{60}$ is about 0.354 nm and is small compared with the wavelength of visible and ultraviolet light, which is supposed to be amplified in the laser resonator. Such scattering is known as Rayleigh scattering. [19] If the incident light is unpolarized with the incident irradiance $I_i$, the scattered irradiance $I_s$ is

$$I_s = \frac{8\pi^4 na^6}{\lambda^4 r^2} \left| \frac{m^2 - 1}{m^2 + 2} \right| (1 + \cos^2\theta) I_i, \qquad (3)$$

where $a$, $\lambda$, and $\theta$ denote the radii of particle, wavelength, and the scattering angle respectively. Besides, the relative refractive index $m = n_1/n$ and $n$ are the refractive indices of particle and medium, respectively) is weakly dependent on wavelength. Because of the angle-dependent $I_s$ in equation (3), the most heavily forward and backward scattering directions, corresponding to $\theta = 0$ and $\pi$, are weighted. Thus the CPPC mirrors act like flat-mirrors, since the forward scattering is forbidden for the wavelength within the photonic gap of CPPCs. Meanwhile, carbon peapods are observed to remain collimating and ordered over distances of 20-50 nm, [15] which assures that the CPPC mirrors separated at the two endcaps of the SWCNT can be well aligned flat-mirrors. Thereby, the nanoscale laser resonator of an effective length of 20-50 nm is stable.

We will consider only the case of a cylindrical empty resonant cavity containing radiation but no matter, as sketched in Figure 2. The assumption that there is radiation but no matter inside the cavity is obviously an approximation if the cavity is part of a working laser. This approximation is used frequently in laser theory, and it is accurate enough for many purpose because laser media are usually only sparsely filled with active atoms or molecules. Since the electric field should vanish at the walls of one-dimensional cavity, we have $\lambda = 2L/n$, where $n = 1,2,\cdots$ is a positive integer, and $L$ the cavity length. This relation indicates that it is impossible for us to have coherent-single-mode-laser emission since the optical cavity dimensions are enormously larger than an optical wavelength ($\lambda \approx 500$ nm). However, the ideally single-mode laser can be easily accomplished by a propitious choice of dimensions of a nanoscale resonator and the wavelengths of the principal (lowest) modes range from 20 to 100 nm. Actually, the nanoscale resonator itself can be constructed on the scale of the wavelengths of interest by synthesizing longer carbon peapods with ordered peas arrays inside. Even the x-ray can be amplified in this kind of nanoscale resonant cavity, if active atoms trapped inside SWCNTS can produce x-ray emission. But the power of the nanoscale laser will be limited in order to remain stability of the resonator.

At the moment, we would like to search for proper active medium in which the extreme-ultraviolet light can be essentially amplified and be conveniently encapsulated inside the carbon peapods. The highest priority should be given to the $C_{60}$ molecules inside the carbon peapods resonator. The fcc crystals are novel semiconductor with direct energy gap of 1.5 eV. [16,20] Note that $C_{60}$ clusters are condensed by van der Waals force, the crystalline phase of $C_{60}$ molecules existing in carbon peapods [10] is strongly suggested to be semiconducting. In spite of the fact that there is no report of laser produced in the semiconducting $C_{60}$ crystals, it deserves detailed examination. An alternative is rare-gas "excimer" (contraction for excited dimer) lasers since their wavelengths tend from the visible to the ultraviolet. For example, the $Ar_2^*$ excimer can produce laser at around 126.1 nm. [21] Meanwhile, these atoms can be trapped inside SWNCTs, [22] and form dimers inside $C_{60}$ and $C_{70}$ fullerenes. [23,24] Hence, they are the most promising active media expected to produce eximer laser in the nanoscale resonant cavity based on carbon peapods.

In summary, we have shown that the carbon peapods with ordered phase of $C_{60}$ molecules inside may be novel one-dimensional *nanostructured* photonic crystals. The calculated photonic bandgap structure of a typical CPPC reveals that x-rays, which have frequencies within the forbidden gap, can be trapped inside the SWCNT. One of the remarkable features of the CPPC is that it is convenient to switch its photonic band gap on and off by voltage-driven changing intermolecular separation of $C_{60}$ molecules inside the SWCNT. Therefore, CPPCs can be employed as optoelectronic circuits in nanodevices. Furthermore, a nanoscale resonant cavity based on CPPCs is supposed to produce *ideally single-mode* laser. Detailed investigations by experiment are desired in the near future.

ACKNOWLEDGMENT. This work is supported by the National Natural Science Foundation of China under Project Nos. 10074053 and 10374080.

**REFERENCES**:

(1) Kroto, H.W.; Heath, J.R.; O'Brien, S.C.; Curl, R.F.; Smally, R.E. *Nature* **1985**, *318*, 162.

(2) Nakanishi, S.; Tsukada, M. *Phys. Rev. Lett.* **2001**, *87*, 126801.

(3) Miura, K.; Kamiya, S.; Sasaki, N. *Phys. Rev. Lett.* **2003**, *90*, 055509.

(4) Shen J.Q.; He, S. *Phys. Rev. B* **2003**, *68*, 195421.

(5) Iijima S.; Ichihashi, T. *Nature* **1993**, *363*, 605.

(6) Smith, B.W.; Mothioux, M.; Luzzi, D.E. *Nature* **1998**, *396*, 323.

(7) Mothioux, M. *Carbon* **2002**, *40*, 1809.

(8) Yakobson B.I.; Smalley, R.E. *Am. Sci.* **1997**, *85*, 324.

(9) Kwon, Y.-K.; Tomanek, D.; Iijima, S. *Phys. Rev. Lett.* **1999**, *82*, 1470.

(10) Khlobystov, A.N.; Britz, D.A.; Ardavan, A.; Briggs, G. A. *Phys. Rev. Lett.* **2004**, *92*, 245507.

(11) Yablonovitch, E. *Phys. Rev. Lett.* **1987**, *58*, 2059.

(12) John, S. *Phys. Rev. Lett.* **1987**, *58*, 2486.

(13) Leung K.M.; Liu, Y.F. *Phys. Rev. Lett.* **1990**, *65*, 2646.

(14) Yeh, P. *Optical Waves in Layered Media*; Wiley: New YorK, 1988.

(15) Smith, B.W.; Russo, R.M.; et al., *J. Appl. Phys. Lett.* **2002**, *91*, 9333.

(16) Saito S.; Oshiyama, A. *Phys. Rev. Lett.* **1991**, *66*, 2637.

(17) Liu, X.; Pichler, T.; Knupfer, M.; Fink, J.; Kataura, H. *Phys. Rev. B* **2004**, *69*, 075417.

(18) Hirahara, K.; Suenaga, K.; Bandow, S.; *et. al.*, *Phys. Rev. Lett.* **2000**, *85*, 5384.

(19) Stratton, J.A. *Electromagnetic Theory*; McGraw-Hill: New York, 1941.

(20) Li, H.; He, S.; Zhang, H.; Lu, B.; Bao S.; Li, H.; He, P.; Xu, Y. *Phys. Rev. B* **2003**, *68*, 165417.

(21) NinomiyaLi H.; Nakamura K. *Opt. Commun.* **1997**, *134*, 521.

(22) Kosmider, M.; Dendzik, Z.; Palucha, S.; Gburski, Z. *J. Mol. Structure* **2004**, *704*, 197.

(23) Baltenkov, A.S.; Dolmatov, V.K.; Manson, *et. al.*, *Phys. Rev. A* **2003**, *68*, 043202.

(24) Ohtsuki, T.; Maruyama, Y.; Masumoto, K. *Phys. Rev. Lett.* **1998**, *81*, 967.